\documentclass[aps,prb,twocolumn,groupedaddresss,showpacs]{revtex4}
\usepackage{amsmath,amsfonts,amssymb,bm,psfrag}
\usepackage{graphics,graphicx}

\renewcommand{\vec}[1]{{\bf #1} }
\newcommand{\RE}{\text{Re}\,}
\newcommand{\IM}{\text{Im}\,}

\newcommand{\eps}{\varepsilon}
\newcommand{\veck}{{\bm k}}

\newcommand{\vecG}{{\bm G}}

\newcommand{\veckappa}{{\bm \kappa}}
\newcommand{\md}{\text{d}}
\newcommand{\me}{\text{e}}
\newcommand{\mi}{\text{i}}

\newcommand{\EF}{\varepsilon_\text{F}}
\newcommand{\kF}{k_\text{F}}
\newcommand{\vF}{v_\text{F}}
\newcommand{\w}{\omega}

\fontencoding{OMS}\fontfamily{cmsy}\selectfont
\begin{document}


\title{Zero temperature optical conductivity of ultra-clean Fermi liquids and
  superconductors}


\author{A. Rosch}
\affiliation{Institut f\"ur Theoretische Physik, Universit\"at zu
  K\"oln, 50937 K\"oln, Germany}
\author{P. C. Howell}
\affiliation{Institut f\"ur Theorie der Kondensierten Materie,
    Universit\"at  Karlsruhe, 76128 Karlsruhe, Germany}

\date{\today}

\begin{abstract}
  We calculate the low-frequency optical conductivity $\sigma(\omega)$ of
  clean metals and superconductors at zero temperature neglecting the effects
  of impurities and phonons. In general, the frequency and temperature
  dependences of $\sigma$ have very little in common.  For small Fermi
  surfaces in three dimensions (but not in $2D$) we find for example that
  $\RE\sigma(\w>0)\approx const.$ which corresponds to a scattering rate
  $\Gamma \propto \w^2$ even in the {\em absence} of Umklapp scattering when
  there is no $T^2$ contribution to $\Gamma$.  In the main part of the paper
  we discuss in detail the optical conductivity of d-wave superconductors in
  $2D$ where $\RE\sigma(\w>0)\propto \w^4$ for the smallest frequencies and the
  Umklapp processes typically set in smoothly above a finite threshold $\w_0$
  smaller than twice the maximal gap $\Delta$. 
In cases where the nodes are located at $(\pm \pi/2,\pm \pi/2)$, such that 
direct Umklapp scattering among them is
  possible, one obtains $\RE\sigma(\w)\propto \w^2$.
  \end{abstract}


\maketitle

{\em Introduction:} Optical conductivity is a powerful tool\cite{dressel} to study the properties
of a strongly correlated metal. The frequency dependence in particular
can give detailed information on the excitation spectrum of a system
(gaps, phonons, magnons, interband transitions, ...) which in
general cannot be extracted from, for example, the temperature
dependence of the conductivity.

In a superconductor the electronic contribution to the optical
conductivity $\RE\, \sigma(\w)$  can be separated -- at least in
simple situations -- into three different contributions. First --
and most important -- superconductivity implies the existence of a
$\delta$-peak at $\w=0$ whose (Drude-) weight is given by the
condensate fraction. Second, thermal excitations at small but finite
temperatures, $T>0$, are expected to lead  to a rather sharp peak
centered again at $\w=0$, whose width is identified with the
scattering rate of the thermal excitations and is strongly
temperature dependent (in cases where impurity scattering can be
neglected). Finally, all other contributions at finite frequency
are usually called ``incoherent background''. This background
depends only weakly on temperature $T$. It is the goal of this paper
to discuss the low-frequency properties of this incoherent
background. More precisely, we consider the optical conductivity for
frequencies $\w>0$ at $T=0$  when thermal excitations are absent.

In most experimentally relevant situations the optical conductivity
of (conventional) superconductors at low frequencies is dominated
by elastic impurity scattering. The theory of optical conductivity in such 
systems was developed very early by Mattis and Bardeen
\cite{mattis}. Inelastic scattering is more important in strongly
interacting superconductors and rather clean samples, and therefore
the optical conductivity in d-wave superconductors has been studied
quite extensively in the context of high-$T_c$ superconductors (see
for example [\onlinecite{orenstein1,carbotte,quinlan}] and
references therein). Motivated by experiments, these investigations have
mainly investigated the influence of scattering from collective
modes and the interplay with impurity scattering. In this paper we
systematically investigate the zero-temperature optical conductivity
arising from the interplay of band-structure effects and
electron--electron interactions taking into account all relevant
vertex corrections. While the main focus of this paper is the
investigation of d-wave superconductors, we also briefly discuss the
optical conductivity of clean Fermi liquids and s-wave
superconductors.

{\em Method:}  According to the Kubo formula, the optical conductivity is
given by
\begin{equation}
  \RE \sigma(\omega) = \frac{1}{\w} \, \IM \left \langle\!\left\langle J , J \right \rangle\!\right\rangle_\w
\end{equation}
where $ \left \langle\!\left\langle J , J \right \rangle
\!\right\rangle_\w$ is the the current--current correlator, $\left
\langle\!\left\langle J , J \right \rangle \!\right\rangle_\w=-\mi
\int_0^\infty
  \md t \,\me^{\mi (\w+\mi 0) t} \langle[J(t),
  J(0)]\rangle.$
When calculating the optical conductivity perturbatively, it is
important to take into account both vertex and self-energy
corrections. For example, in a Galilean invariant system with a
quadratic dispersion, $\eps_k = k^2/(2 m)$, vertex and
self-energy corrections cancel exactly as the total electrical
current is a conserved quantity. But even in clean non-Galilean
invariant systems, i.e. for electrons moving in a periodic
crystalline potential, massive cancellations between self-energy and
vertex corrections occur, especially if there is little Umklapp
scattering close to the Fermi surface.  To take into account vertex
and self-energy corrections on the same footing, one in general has 
to solve an integral equation (a vertex equation or, equivalently, a
linearized quantum Boltzmann equation) to obtain the correct
conductivity even to lowest order in perturbation theory.

However, at zero temperature and in the absence of disorder one can
avoid the substantial technical difficulties involved in solving
multi-dimensional integral equations by the following argument: In
general, one can express the conductivity in the form
$\sigma(\w)=\frac{\chi}{\Gamma(\w)-i \omega}$ where $\chi$ is identified with
the total optical weight and the
(frequency-dependent) scattering rate $\RE \Gamma(\w)$ can be
calculated from the integral equations described above. However, for
$ |\Gamma(\w)| \ll \w$ this simplifies after multiplication with
$\w^2$ to
\begin{equation}\label{sigG}
\w^2 \RE\, \sigma(\w)=\w^2 \, \RE\, \frac{\chi }{\Gamma(\w)-i
\omega} \approx \chi \RE\,\Gamma(\w)
\end{equation}
Note that there is no contribution from the $\delta$-function at
$\w=0$ due to the $\w^2$ prefactor. For weak interactions $\Gamma$ is small
and therefore we can obtain $\sigma(\w>0)$ in a
straightforward perturbative expansion, i.e. without solving any
integral equations, from the right-hand side of
\begin{eqnarray}\label{sigPT}
\RE \, \sigma(\w>0) =  \frac{\IM \left \langle\!\!\left\langle
\partial_t J,
\partial_t J \right \rangle \!\!\right\rangle_\w}{\w^3},
\end{eqnarray}
provided that $ |\Gamma(\w)| \ll \w$.  As $\partial_t J$ is already linear in
the interactions (see below), it is sufficient to leading order to evaluate
the correlation function in (\ref{sigPT}) to zeroth order in the couplings. We
will use this approximation only at $T=0$. At
any {\em finite} temperature, the scattering rate $\Gamma(\w\to 0)$ is constant and
therefore the method described above will break down for $\w \to 0$  but
remains valid at higher frequencies where $|\Gamma(\w)|\ll \w$.  Note that
within the so-called memory-function approach\cite{memory} one uses
essentially identical formulas to calculate $\Gamma(\w)$.

  If one is only interested in the qualitative behavior of $\RE \sigma(\w)$ at
  low $\w$, i.e. in the power-law obtained for $\w \to 0$, one can relax the
  condition $|\Gamma(\w)|\ll \w$ and replace it by $\RE \Gamma(\w)\ll (1-c)
  \w$ for $\w<\w^*$ where $\w^*$ is some characteristic frequency and $c$
  (which can be of order $1$) is obtained from $\IM \Gamma(\w)\approx c \w$
  for $\w\ll w^*$.  As the latter condition is fulfilled in all cases
  discussed below, we expect that all our results are {\em qualitatively}
  correct at sufficiently low frequencies (a possible exception is discussed
  below) even in a strongly interacting systems.  (Backflow and other Fermi
  liquid renormalization effects\cite{fermiLiquid} will only change prefactors,
  and multi-particle scattering processes are suppressed for $\w \to 0$ due to
  the restricted phase space.)

In the following, we will first consider the optical conductivity of
a clean Fermi liquid at $T=0$. This will serve as a reference for
our results on d-wave superconductors presented in the second part.

{\em Metals:} In a one-band model, the electrical current is given by
$\vec{J}=\sum_{\vec{k}\sigma} \vec{v}_{\vec{k}}
c_{\vec{k}\sigma}^\dagger c_{\vec{k}\sigma}$. In the presence of
interactions and in the absence of Galilei invariance the current is
not conserved with $\partial_t \vec{J}=\mi \sum_{\vec{k}
\vec{k}'\vec{q}\sigma\sigma'}U_\vec{q}
(\vec{v}_{\vec{k}}+\vec{v}_{\vec{k}'}-\vec{v}_{\vec{k}+\vec{q}}-
\vec{v}_{\vec{k}'-\vec{q}})c_{\vec{k}\sigma}^\dagger
c_{\vec{k}+\vec{q},\sigma}c_{\vec{k}'\sigma'}^\dagger
c_{\vec{k}'-\vec{q},\sigma'} $ for a density--density interaction
$U_\vec{q}$. The change of current is proportional to the difference
of incoming and outgoing velocities. In the following we will assume
that the (screened) interaction $U_\vec{q}\approx U$ depends only weakly
on the transferred momentum $\vec{q}$.

For a (normal) metal we therefore obtain at low frequencies and $T=0$ using Eq.~(\ref{sigPT})
\begin{multline}\label{sigFL}
\RE\,\sigma(\omega>0) \approx \frac{4 \pi U^2}{\w^3} \sum_{1234\vecG}
f_1 f_2 (1-f_3)(1-f_4)\\
  \times   (v_4^x+v_3^x-v_2^x-v_1^x)^2 \, \delta_{1+2,3+4+\vecG}\\
\times \left[\delta\!\left(\omega-
(\eps_4+\eps_3-\eps_2-\eps_1)\right)-(\w \leftrightarrow -\w)\right]
\end{multline}
where $1,...,4$ denote the momenta $\vec{k}_1,...,\vec{k}_4$ in the
first  Brillouin zone and momentum is conserved modulo reciprocal
lattice vectors $\vec{G}$. To perform the momentum integrals it is
useful to split $\vec{k}_i$ into a component perpendicular to the
Fermi surface and an angular integration parallel to it. For small
$\w$ only a thin shell of width $\w/v_F$ contributes for each of the three
relevant momentum integrations, implying an $\w^3$
dependence which cancels the $1/\w^3$ prefactor.

If the Fermi surface is sufficiently large such that Umklapp
scattering processes can take place, one therefore obtains the 
well-known result that
\begin{equation}\label{flU}
\RE\,\sigma(\omega>0) \approx const., \qquad \Gamma(\w)\propto \w^2
\end{equation} as 
the 4 velocities sum up to a finite value of the order of $v_F$ in this case.
The constant ``incoherent background'' corresponds according to
Eq.~(\ref{sigG}) to a scattering rate $\Gamma(\w) \propto \w^2$ characteristic
of a Fermi liquid with Umklapp scattering in 2 or 3 dimensions.

Less well known is the corresponding result for a small Fermi
surface ($k_F<G/4$) where Umklapp scattering at the Fermi surface is
not possible. Here the situations in 2 and 3 dimensions are quite
different. For a generic (not too complex) Fermi surface in 2$D$,
momentum conservation in the limit $\w\to 0$ can only be fulfilled
by choosing $\vec{k}_1=-\vec{k}_2$ and $\vec{k}_3=-\vec{k}_4$ (or
$\vec{k}_1=\vec{k}_{3/4}$ and $\vec{k}_2=\vec{k}_{4/3}$). Therefore
the sum of the velocities also vanishes linearly in $\w$ for $\w \to 0$ and one
obtains from power counting
\begin{equation}
\RE\,\sigma(\omega>0) \propto \w^2, \qquad \Gamma(\w)\propto \w^4  \label{cond2d}
\end{equation}
 for a small Fermi surface in $d=2$.

The situation is quite different for a system with a small Fermi
surface in $3D$, where momentum conservation on the Fermi surface
does {\em not} require that the relevant moments are located
opposite to each other. Therefore the sum of the 4 velocities in
Eq.~(\ref{sigFL}) will generically {\em not} vanish  and one finds
\begin{equation}
\RE\,\sigma(\omega>0) \approx const., \qquad \Gamma(\w)\propto \w^2  \label{cond3d}
\end{equation}
for a small Fermi surface in $d=3$: Even {\em without} Umklapp
processes the scattering rate varies as $\Gamma(\w)\propto \w^2$!
Note that the frequency and temperature dependence of the
conductivity are drastically different in this case. $\Gamma(T)$
does {\em not} vary as $T^2$ but the two-particle scattering rate is
exponentially suppressed; multi-particle processes lead to a
power law $\Gamma \propto T^{2 n-2}$ where the integer $n$ depends on
the size of the Fermi surface\cite{roschFL}, $n \sim G/(2 k_F)$.
The disparate behavior of $\Gamma(\w,T=0)\sim U^2 \w^2$ and
$\Gamma(\w=0,T)\sim U^2 \me^{-\Delta/T}+U^{n} T^{2 n-2} $ can easily
be understood once one realizes that,  in
the absence of Umklapp scattering, on the one hand the current is
not conserved while on the other hand the momentum is conserved. 
As explained in detail e.g. in
Refs.~[\onlinecite{roschPRL,roschFL}], the component of the current
``perpendicular'' to the momentum does decay rapidly giving rise to the
frequency independent incoherent background of Eq.~(\ref{cond3d}).
The dc conductivity is, however, determined by the long-time decay
of the component of the current ``parallel'' to the momentum and
therefore by the decay rate of the momentum, i.e. by Umklapp
processes which are very rare for small Fermi surfaces. It is likely
that the rather general results Eq.~(\ref{cond2d},\ref{cond3d}) have
been discussed before in the literature but we are not aware of a
directly relevant reference.

It should be clear from the discussion of Eq.~(\ref{cond3d}) given above that
also in the {\em presence} of Umklapp scattering, when $\Gamma(\w,T)\approx a
(k_B T)^2+ b (\hbar \w)^2$ there is in general no simple relation between the
constants $a$ and $b$. We emphasize this fact as in the experimental
literature such a relation has sometimes been claimed to
exist\cite{dressel,reviews,exp1} but is actually not observed\cite{exp1,scheffler}.
Note that recent progress in the experimental methods allows precise
measurements of the optical conductivity at low frequencies and
temperatures\cite{scheffler}.

{\em Superconductors:} We now turn to the calculation of the $T=0$ optical
conductivity in superconductors neglecting again phonons and impurities. Our
main interest is  the case of a d-wave
superconductor in $d=2$ on a square lattice with 
unit lattice spacing.
To describe the superconducting state we use weakly-interacting Bogoliubov
quasiparticles (QPs), $d^\dagger_{\veck\sigma} = u_\veck
c^\dagger_{\veck\sigma} - \sigma v_\veck c_{-\veck\bar{\sigma}}$, which diagonalize the BCS Hamiltonian $H_{\text{BCS}}=\sum_{\vec{k}\sigma} \eps_\veck c^\dagger_{\veck\sigma}
c_{\veck\sigma}+\sum_{\vec{k} } \Delta^*_\vec{k} c^\dagger_{\veck\uparrow}
c^\dagger_{-\veck\downarrow}+h.c.=\sum_{\vec{k}\sigma} E_\veck
d^\dagger_{\veck\sigma}
d_{\veck\sigma}$
where $c^\dagger_{\veck\sigma}$ is the electron creation operator and $E_\veck = \sqrt{\eps_\veck^2 +
  \Delta_\veck^2}$  the BCS energy. The electric current is given by
\begin{equation}
  \vec{J} = \sum_{\veck\sigma}\vec{v}_\veck
  c_{\veck\sigma}^\dagger c_{\veck\sigma}
  = \sum_{\veck\sigma}\vec{v}_\veck
  d_{\veck\sigma}^\dagger d_{\veck\sigma}
  \;,
\end{equation}
where it is important to realize that it is the {\em bare} velocities
$\vec{v}_{\vec{k}}=\md \eps_\vec{k}/\md \vec{k}$ rather than 
$\md E_\vec{k}/\md \vec{k}$ that enter if the current is expressed in terms 
of the BCS quasiparticles.

Within the BCS approximation the current is conserved, $[J,H_{BCS}]=0$, and there is no optical
weight at finite frequencies. To calculate the optical conductivity
it is therefore essential to include the interaction of the
quasiparticles. The Hamiltonian for the QPs is given by ${\cal H} =
{H}_{\text{BCS}} + {H}_\text{int}$, where the (properly normal
ordered) local density-density interaction $H_{\text{int}}=2 U \sum_i n_{i\downarrow}
n_{i\uparrow}$ can be rewritten as
\begin{widetext}
\begin{multline}\label{HH}
{H}_\text{int} = U \sum
  r_{13} r_{24}
  d^\dag_{4\uparrow} d^\dag_{3\downarrow} d^\dag_{2\downarrow}
  d^\dag_{1\uparrow} +h.c. + 2 \tilde{r}_{12} r_{34}
  d^\dag_{4\uparrow} d^\dag_{3\downarrow} d^\dag_{2\sigma} d_{1\sigma}+h.c.
  + r_{12} r_{34}
  d^\dag_{4\sigma} d^\dag_{3\bar{\sigma}} d_{2\bar{\sigma}}d_{1\sigma}
 +  \tilde{r}_{14} \tilde{r}_{23}
  d^\dag_{4\sigma} d^\dag_{3\sigma'} d_{2\sigma'} d_{1\sigma}
\end{multline}
where   $r_{ij}=r_{ji} = u_1v_2 + v_1u_2$, $\tilde{r}_{ij}
=\tilde{r}_{ji} = u_1u_2 - v_1v_2$ and $i \equiv \bm{k}_i$. The
momentum sums conserve crystal momentum and the spin sums are {\em
only} over repeated indices. This expression can be derived by
keeping the fluctuations around mean-field theory in the BCS
approach. The various terms describe not only the scattering of
quasi-particles (and holes) but also the breaking up and
recombination of Cooper pairs.

While $[J,{H}_\text{BCS}]=0$, the current $J$
decays in the presence of  the interactions between QPs, $\partial_t J=-\mi [J,H_\text{int}]$.
It is now straightforward (albeit somewhat tedious) to evaluate the
contributions to (\ref{sigPT}) to lowest order in the interactions,
and we obtain
\begin{align}\label{all}
  \RE\, \sigma(\omega) &  = \frac{\pi U^2}{\w^3} \left(
 \phi''_\text{pp}(\omega) + \phi''_\text{pq}(\omega)
    + \phi''_\text{qq}(\omega)- ( \omega \leftrightarrow -\omega) \right) \\
  \phi''_\text{pp}(\omega) &= \sum_{1234\vecG}
    (r_{12}r_{34} - r_{13}r_{24})^2
    \,\delta_{1+2+3+4-\vecG} \,(v^x_1+v^x_2+v^x_3+v^x_4)^2 \notag \\
  & \quad\times \left[(1-f_1)(1-f_2)(1-f_3)(1-f_4)-f_1f_2f_3f_4\right]
    \delta(\omega-(E_1+E_2+E_3+E_4))  \notag \\
  \phi''_\text{pq}(\omega) &=4 \sum_{1234\vecG}
    (\tilde{r}_{12}r_{34}-\tilde{r}_{14}r_{23})^2
    \,(v^x_1-v^x_2-v^x_3-v^x_4)^2\, \delta_{1-2-3-4+\vecG} \notag \\
  & \quad\times
\left[f_1(1-f_2)(1-f_3)(1-f_4)-(1-f_1)f_2f_3f_4\right]
    \delta(\omega-(-E_1+E_2+E_3+E_4))  \notag \\
  \phi''_\text{qq}(\omega) &= \sum_{1234\vecG} (1-f_1)(1-f_2)f_3f_4\,
    (v^x_4+v^x_3-v^x_2-v^x_1)^2\,\delta_{1+2-3-4+\vecG} \notag \\
  & \quad\times\left[4 ( r_{12} r_{34}+  \tilde{r}_{14}\tilde{r}_{23})^2
    + 2 (\tilde{r}_{14}\tilde{r}_{23}
    - \tilde{r}_{13}\tilde{r}_{24})^2 \right]
    \delta(\omega-(E_1+E_2-E_3-E_4))  \notag
\end{align}
\end{widetext}
The first (second, third) contribution comes from the first (second, last two)
scattering terms in the Hamiltonian~(\ref{HH}). At zero temperature, obviously
only the first term $\phi_\text{pp}(\omega)$ survives as all QPs have positive
energies and $f_i=f(E_{\vec{k}_i})=0$ at $T=0$.

In the case of an ultra-clean s-wave superconductor, a direct consequence of
Eq.~(\ref{all}) is that the gap in the optical conductivity is of size $4
\Delta$ while it is $2 \Delta$ for dirty superconductors\cite{mattis} as has
previously been noted by Orenstein {\it et al.}\cite{orenstein1} Obviously one
has to ask whether this result will also hold to higher order in perturbation
theory. To answer this question, one has to investigate whether symmetries and
corresponding selection rules allow for an optical transition from the
ground state of the superconductor to a 2-quasiparticle excited state by an
operator of the form $\sum_{\vec{k} \vec{k}'}
\alpha^{\sigma\sigma'}_{\vec{k}\vec{k}'} \tilde{d}_{\vec{k} \sigma}^\dagger
\tilde{d}_{-\vec{k}' \sigma'}^\dagger$ where $\tilde{d}^\dagger$ are the
creation operators of the fully renormalized ``true'' quasiparticles of the
system (which can only be identified with the BCS quasiparticles for weak
interactions). Symmetries strongly restrict the form of
$\alpha^{\sigma\sigma'}_{\vec{k}\vec{k}'}$. Translational invariance on the
lattice, for example, implies that
$\alpha^{\sigma\sigma'}_{\vec{k}\vec{k}'}=\alpha^{\sigma\sigma'}_{\vec{k}}
\delta(\vec{k}-\vec{k}')$ in the absence of impurities with
$\alpha_{\vec{k}}^{\sigma \sigma'}=-\alpha_{-\vec{k}}^{\sigma' \sigma}$ as the
quasiparticles are fermions.  If the superconductor does not break
time-reversal invariance one has $(\alpha_{\vec{k}}^{\uparrow
  \uparrow})^*=(\alpha_{\vec{k}}^{\downarrow \downarrow})$ and
$(\alpha_{\vec{k}}^{\uparrow \downarrow })^*=-(\alpha_{\vec{k}}^{\uparrow
  \downarrow})$ and in a crystal with inversion symmetry one has
$\alpha^{\sigma\sigma'}_{\vec{k}}=\alpha^{\sigma'\sigma}_{\vec{k}}$.  In the
absence of spin-orbit coupling, i.e. if spins are rotationally invariant one
finds that $\alpha_{\vec{k}}^{\uparrow \downarrow}=
-\alpha_{\vec{k}}^{\downarrow \uparrow}$ and
$\alpha^{\sigma\sigma}_{\vec{k}}=0$. From this we can conclude that, in the
absence of disorder and in the presence of inversion symmetry,
$\alpha_{\vec{k}}^{\sigma \sigma'}$ vanishes and the optical gap is therefore
$4\Delta$ for an s-wave superconductor in the {\em absence} of spin-orbit
coupling. In the presence of impurities, however, the gap\cite{mattis} is only
$2 \Delta$. Interestingly, the symmetry analysis suggests that even in a
generic inversion-symmetric clean crystal, high-order spin-orbit processes
could possibly induce relevant low-energy processes not included in
(\ref{all}) which lead to a gap of size $2 \Delta$. All the low-order results
presented below may therefore not be valid in the presence of sizable
spin-orbit coupling. Note also that phonons and other low-energy collective
modes with energies smaller than $2 \Delta$ can induce optical weight in the
frequency window $2 \Delta < \w < 4 \Delta$.  The precise functional form of
the optical conductivity of an s-wave superconductor for $\w \gtrsim 4 \Delta$
will not be discussed in detail here. It depends on the dimension and on the
angular dependence of $\Delta$. Generically the onset will be smooth and of
the form $(\w-4 \Delta)^2$. Therefore a precise experimental determination of
$\Delta$ using a feature close to $4\Delta$ will be rather difficult. For all
conventional s-wave superconductors we anyhow expect that impurity scattering
will dominate even for the cleanest available samples leading to the
well-known $2 \Delta$ gap which is much easier to detect.

A d-wave superconductor in two dimensions (as realized in high-temperature
superconductors) with point nodes along the diagonals
of the quadratic Brillouin zone has a vanishing gap in nodal direction.
For frequencies small compared to the maximal gap $\Delta$, $\w \ll\Delta$, all
QPs are created in the vicinity of the nodes, so we expand the dispersion
around them. Writing $\veck = \veck_\text{node} + \veckappa$ the most generic
band structure consistent with the square symmetry of the lattice is
\begin{align}\label{disp}
  \eps_\veckappa &=
    \frac{v_F}{\sqrt{2}} (\kappa_x+\kappa_y) +\frac{1}{2m^*} (\kappa_x^2+\kappa_y^2)
    + D\kappa_x\kappa_y \notag \\
  & \quad +L(\kappa_x^3+\kappa_y^3) + F\kappa_x\kappa_y(\kappa_x+\kappa_y)
  + {\cal O}(\kappa^4)
\end{align}
where the constants $D,L,F$ determine the deviation of the
dispersion from that of a free-electron gas and $m^*$ is an
effective mass.

\begin{figure}
\includegraphics[width=80mm]{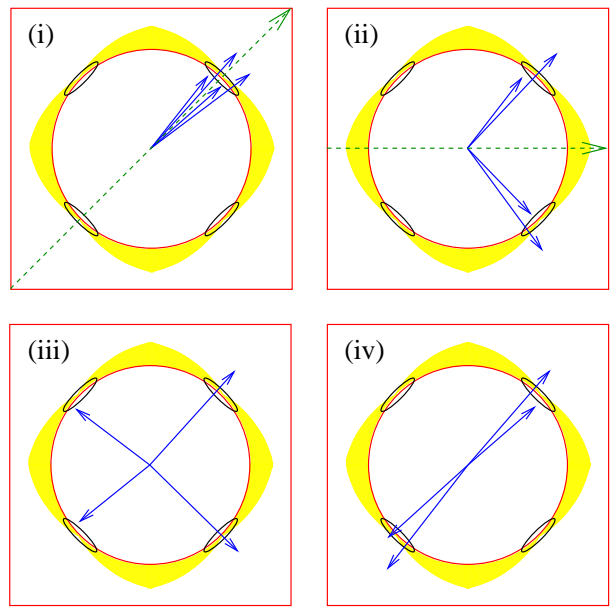}
\caption{\label{Fig:zeroTscatposs}
  The four possible types of scattering process at $T=0$, in which
  four QPs are created. The circle represents the Fermi surface, the
  solid arrows the QP momenta, and the dashed arrows the reciprocal
  lattice vector $\vecG$. The shaded region indicates the size of the
  superconducting gap and the ellipses a constant energy contour in
  the vicinity of each node.  }
\end{figure}

There are four qualitatively different terms that appear in the sum
 for $\phi''_\text{pp}$ in (\ref{all}), which are sketched in Fig.~\ref{Fig:zeroTscatposs}:
(i)~$\vecG=(2\pi, 2\pi)$ and hence all
four QPs in one node; (ii)~$\vecG=(2\pi,0)$ and
two QPs in each of two  ``perpendicular'' nodes; (iii)~$\vecG={\bm
0}$ and one QP in each node; (iv)~$\vecG ={\bm 0}$ and two QPs in
each of two opposite nodes. These give rise to very different
dependences on $\omega$ and doping, as we now discuss.

The role of Umklapp scattering is determined by the distance of the nodes
from $(\pi/2,\pi/2)$ which we denote by
\begin{equation}
\delta k_\text{node} = \left|\vec{k}_\text{node}-\left(\!\frac{\pi}{2}
    ,\frac{\pi}{2}\!\right)\right|=\left|k_F  -\frac{\pi}{\sqrt{2}}\right|
\end{equation}
In processes of type (i) the four QPs have very similar velocities
(recall that it is the {\em normal-state} velocity that contributes)
and so the contribution to $\phi''_\text{pp}(\omega)$ is large.
However, it is only possible to create four QPs of arbitrarily low
energy if the nodes are situated exactly at
$(\pi/2,\pi/2)$; otherwise there is an excess
momentum $4 \delta k_\text{node}$ which must be carried by the QPs,
so that at least one of them is situated a finite distance from the
node. Accordingly, absorption can only occur for frequencies above
the threshold \begin{equation} \label{w0}
 \omega_0 \approx 4 v_\text{F} \delta
k_\text{node}.\end{equation}

Processes of type (ii) resemble those of type (i), since again the
velocities add. However, the fact that the nodes are at right angles
to one another reduces the threshold frequency
 as the excess momentum $\big(\sum\veck_\text{node}\big) - \vecG =
(2\sqrt{2} \delta k_\text{node},0)=\sqrt{2}\delta k_\text{node}
[(1,1)+(1,-1)]$ can be split into two components parallel to the Fermi surface at
the nodes where the velocity of the QPs $v_\Delta=\md
E_k/\md k_\parallel=\md \Delta_k/\md k_\parallel$ is much smaller. This leads to a
considerably smaller threshold frequency
\begin{equation} \label{w0s}
\omega_0'
\approx 4 v_\Delta \delta k_\text{node} \sim
\frac{\Delta}{\EF} \omega_0\;, \end{equation} 
where this
simplified formula is only valid if $\delta k_\text{node} <
\Delta/\vF$ when corrections to the Dirac spectrum close to the
nodes can be neglected. Note that the construction described above
reduces the available phase space for scattering and so the
contribution close to $\w_0'$ is smaller than that of type (i)
processes by a factor of $v_\Delta/\vF$.

\begin{figure}
  \includegraphics[width=0.98 \linewidth]{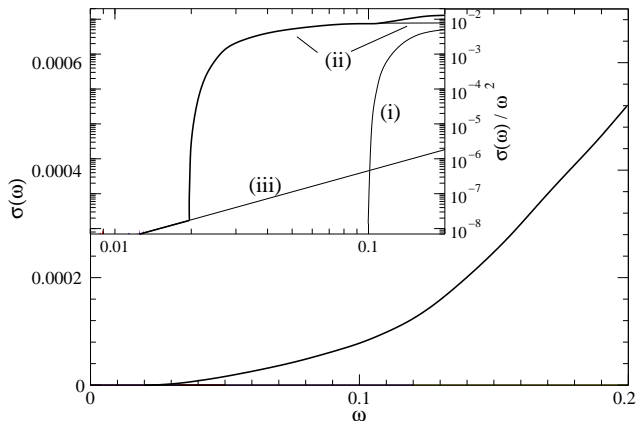}
\caption{\label{toy}  Zero-temperature
  optical conductivity, $\RE \sigma(\w>0)$, for an ultra-clean d-wave
  superconductor from a numerical evaluation of Eq.~(\ref{all}) for a model
  with nodes close to
  $(\pi/2,\pi/2$) [$\Delta=0.31, v_F=1, v_F/v_\Delta=5, \w_0=0.1$ and
  $\w_0'=0.02$].  Inset: logarithmic plot of $\RE \sigma(\w>0)/\w^2$. The thin
  lines show the contributions from the various processes shown in
  Fig.~\ref{Fig:zeroTscatposs}.  Processes of type (i) and (ii) are gapped by
  $\w_0$ and $\w_0'$, respectively.  Due to the smooth onset of type-(i)
  Umklapp processes, Eq.~(\ref{sig1gap}), there is almost no feature at $\w_0$.
  The numerical results are fully consistent with the power-laws of 
Eqs.~(\ref{sig1}--\ref{sig2gap}).  }
\end{figure}

\begin{figure}
\includegraphics[width=\linewidth]{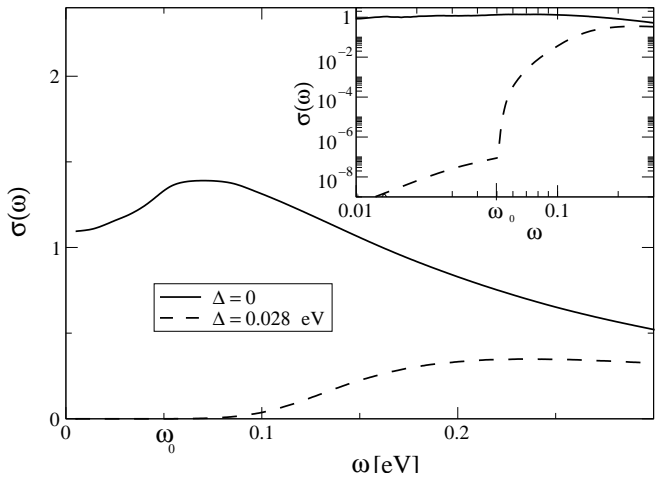}
\caption{\label{real} Optical conductivity, $\RE \sigma(\w>0)$, for an ultra-clean d-wave
  superconductor (dashed line) using the bandstructure of optimally doped
  Bi-2212 taken from [\onlinecite{bandstructure}] and a d-wave gap of size
  $\Delta=0.028$\,eV (arbitrary units on the y-axis). For reference, the solid
  line shows the $T=0$ optical conductivity in the normal state ($\Delta=0$)
  which is constant for low frequencies due to Umklapp processes, see
  Eq.~(\ref{flU}). The $\delta$-peak at $\w=0$ is not shown. For lowest
  frequencies, $\w < \w_0$, one finds $\sigma \propto \w^4$ in the
  superconducting phase as can be seen on the logarithmic scale of the inset.
  This regime is, however, practically not observable due to the small
  prefactor. Instead one finds a very smooth onset [see Eq.~(\ref{sig2gap})]
  for $\w>\w_0\approx 0.05\,\text{eV} < 2 \Delta$ (\ref{w0ss}).  Note that the
  nodes are {\em not} close to $(\pi/2,\pi/2)$ for the bandstructure
  considered in this figure.}
\end{figure}

In most realistic situations (including most of the cuprates) the
point node will {\em not} be located close to $(\pi/2,\pi/2)$ and
$\delta k_\text{node}$ will be larger than $\Delta/v_F$. In this
case the gap for Umklapp processes will depend on details of the
band-structure. For sufficiently large Fermi surfaces  (e.g.
optimally doped Bi-2212 according to
Ref.~[\onlinecite{bandstructure}], see also Fig.~\ref{real}), the gap for Umklapp processes in
a d-wave superconductor will be smaller than $2 \Delta$,
\begin{equation} \label{w0ss} \w_0 < 2 \Delta
\end{equation}  as typically an Umklapp process will exist where two
QPs are located at the nodes and the two other somewhere else
on the Fermi surface. 

To obtain the frequency dependence, we ignore in a first step the coherence
  factors and velocity prefactors and evaluate the integral
\begin{equation}\label{s}
\sum_{1,2,3}
    \delta(\omega-(E_1+E_2+E_3+E_{-(1+2+3)}))\approx 
    \frac{c_i \w^5}{(v_F v_\Delta)^3}
\end{equation}
for $\delta k_\text{node}=0$ and small $\w$
 in each of the 4 cases shown in
Fig.~\ref{Fig:zeroTscatposs}. This can be done by scaling the
momenta perpendicular and parallel to the Fermi surface at the node by
$\w/v_F$ and $\w/v_\Delta$, respectively. In the cases (i) and (iv)
shown in Fig.~\ref{Fig:zeroTscatposs}, when all nodes are parallel
to each other, $c_1$ and $c_4$ are constants of order $1$. The
situation is slightly more complicated in the cases (ii) and (iii)
where by choosing a proper rescaling procedure we find $c_2\sim
c_3\sim v_\Delta/v_F$.

Eq.~(\ref{s}) does not include the effect of the velocity prefactor $(\sum
v^x_i)^2$ and of the combination $(r_{12} r_{34}-r_{13} r_{24})^2$ of coherence factors in
Eq.~(\ref{all}). At the nodes, the coherence factors $\frac{1}{\sqrt{2}}
\sqrt{1 \pm \frac{\eps_\vec{k}}{E_{\vec{k}}}}$ are rapidly varying
functions of order 1. For $\delta k_\text{node}=0$ they change the result only
quantitatively but not qualitatively (as we have checked numerically) but can become
important for $\delta k_\text{node}\neq 0$ as discussed below.  For the
Umklapp processes (i) and (ii) the velocities just add up to give a finite
prefactor of order $v_F^2$. If the nodes are located at $(\pi/2,\pi/2)$, we
therefore obtain
\begin{equation}\label{sig1}
\sigma(\w)\propto \frac{U^2}{v_F v_\Delta^3}\, \w^2, \qquad
\text{type (i) for } \delta k_\text{node}=0
\end{equation}
and similarly
\begin{equation}\label{sig2}
\sigma(\w)\propto \frac{U^2}{v_F^2 v_\Delta^2}\, \w^2, \qquad
\text{type (ii) for } \delta k_\text{node}=0.
\end{equation}
These power-laws can also be observed for $\w \gg \w_0,\w_0'$ if  $\Delta
k_\text{node}$ is finite but small. This can be seen in the inset of
Fig.~\ref{toy}) which discusses the various regimes based on a  numerical
evaluation of  Eq.~(\ref{all}).

Due to momentum conservation, the leading  contribution to $\sum
v^x_i$ vanishes for the non-Umklapp processes (iii) and (iv). But
band-structure effects break Galilean invariance and one obtains a
low-energy contribution even in the absence of Umklapp. The leading
term is given by $D\kappa_x\kappa_y$ in  (\ref{disp}) which leads to 
$v_x = D \kappa_y$. Although the sum $\sum_i v_i^x$ still vanishes at this order
for processes of type (iv), it remains finite if the geometry is
determined by (iii) and we obtain from a scaling analysis
\begin{equation} \label{sig3}
\sigma(\w)\propto \frac{U^2 D^2}{v_F^6 v_\Delta^2} \, \w^4
\hspace{1cm} \text{type (iii) for } \, \w \to 0.
\end{equation}
While this term is suppressed by the tiny factor $\frac{\w^2
  v_\Delta}{\EF^2 v_F}$  compared to (\ref{sig1}), it is nevertheless
the leading $\w \to 0$ correction when the nodes are located {\em
away} from $(\pi/2,\pi/2)$. Eq.~(\ref{sig3}) therefore describes the
typical low-frequency optical conductivity of a 2-dimensional d-wave
superconductor in the absence of impurities (c.f. insets of Figs.~\ref{toy},\ref{real}). Processes from the
scattering geometry (iv) are always subleading and only give rise to
contributions $\propto \w^6$. It is worth noting that the
prefactor of
  (\ref{sig3}) -- not shown in the equation -- turns out to be numerically
  very small, approximately a factor
  of $20$ smaller than the prefactor of (\ref{sig1}) and more than a factor
  of $100$ smaller than the corresponding numerical prefactor of  (\ref{sig2}) if we assume a
  local interaction $U$. In general completely different matrix elements enter
  the various scattering processes (i) -- (iv) and therefore their relative
  magnitude depends on details of the relevant interactions.
But the smallness of the contribution may imply that in actual measurements the low-frequency $\w^4$ regime is
  never observable, see Figs.~\ref{toy} and \ref{real}.

As the non-Umklapp contribution (\ref{sig3}) to the optical
conductivity is very  small and difficult to detect experimentally,
it is worthwhile to investigate the precise form of the onset of
Umklapp terms at $\w>\w_0,\w_0'$ in the generic case when the nodes
are not located at $(\pi/2,\pi/2)$. Consider for example the
scattering geometry (i) in Fig.~\ref{Fig:zeroTscatposs}. At
$\w=\w_0$ the  components $\kappa_\|$ of all 4 momenta {\em parallel} to the Fermi surface
will be zero, 
so $\eps_\vec{k} = E_{k}$ and therefore the coherence
factors $(r_{12} r_{34}-r_{13} r_{24})^2$ of Eq.~(\ref{all}) will
vanish. As a consequence the onset of Umklapp processes will be very
smooth and of the form
\begin{equation}\label{sig1gap}
\sigma(\w)\propto \frac{U^2}{v_F v_\Delta^3} \, (\w-\w_0)^2, \qquad
\text{type (i) for } \w \gtrsim \w_0
\end{equation}
and
\begin{equation}\label{sig2gap}
\sigma(\w)\propto \frac{U^2}{v_F^2 v_\Delta^2} \, (\w-\w_0')^2,
\qquad \text{type (ii) for } \w \gtrsim \w_0'
\end{equation}
as we have checked numerically, see Figs.~\ref{toy} and \ref{real}. Formulas for $\w_0$ and $\w_0'$ are given
in Eqs.~(\ref{w0}--\ref{w0ss}) above. The prefactors in
Eqs.~(\ref{sig1gap},\ref{sig2gap}) are only valid for very small
$\w_0,\w_0'\ll \kF v_\Delta^2/\vF$ when one can use a Dirac spectrum
for the nodal quasiparticles; however, the frequency dependence close to the
onset frequency is also quadratic for larger values of
$\w_0$ and $\w_0'$ as we have again checked numerically for example in
Fig.~\ref{real} which shows the optical conductivity in a model which uses the
bandstructure\cite{bandstructure} of Bi-2212.

All results shown above rely on the fact that at lowest energies the
nodal dispersion takes the form of a Dirac cone,
$E_\vec{k}=\sqrt{(v_F k_\perp)^2+(v_\Delta k_\|)^2}$. But already at
a very low energy scale, $E_c= m^* v_\Delta^2/2 \sim
\Delta^2/\EF \ll \Delta$, one has to take into account the
curvature of the Fermi surface which bends contours of equal energy
into a banana shape. It is therefore important to check which of the
results calculated above remain unmodified at this crossover scale
-- the existence of such a small energy scale will otherwise make
the experimental determination of power laws extremely difficult.
Fortunately, it turns out that our results in the scattering
geometry (ii) and (iii), i.e. Eqs.~(\ref{sig2},\ref{sig3}), are not
affected by $E_c$ and remain valid up to energies of the order of
the maximal gap $\Delta$. This can most easily be seen by rewriting
momentum conservation in polar coordinates while scaling  $k-k_F$
with $\w/v_F$ and the polar angle $\phi$ with $\w/(k_F v_\Delta)$.
Using the same analysis for geometry (i), one finds a crossover at
the energy $E_c$ and Eq.~(\ref{sig1})
 has to be multiplied by a
factor $E_c/\w$ for $E_c \ll \w \ll \Delta$.

{\em Conclusions:} The frequency dependence of the optical conductivity at
zero temperature and finite frequencies describes how the electrical current
can decay. The example of a Fermi liquid with a small Fermi surface shows that
the temperature and frequency dependencies of $\sigma(\w,T)$ have very little
in common and may result from completely different processes. The
zero-temperature optical conductivity of d-wave superconductors turns out to be
rather complex even for frequencies much smaller than the maximal gap. While we
hope that our calculation can serve as a reference for the interpretation of
the incoherent background, a direct observation of the predicted power-laws
will be difficult as the calculated contributions turn out to be both small in
size and very smooth in their frequency dependence (see Fig.~\ref{toy} and \ref{real}). Therefore it will be very
difficult even in very clean crystals to separate the predicted effects from
the effects of elastic impurity scattering.

An interesting open question is whether and how spin-orbit interactions modify
the results presented in this paper. Based on a symmetry analysis, we argued
that spin-orbit interactions can open new channels for current relaxation in a
superconductor -- even in the presence of inversion symmetry. Neglecting such
relativistic effects, we believe that our results are valid even in strongly
interacting superconductors at sufficiently low frequencies when multi-particle
scattering is suppressed due to phase-space restrictions. This will also be
the case if the interactions are mediated e.g. by (short-ranged)
spin-fluctuations\cite{carbotte,quinlan}, provided the system is not located
directly at a quantum-critical point.

At small but finite temperatures thermal excitations induce a characteristic
sharp peak in the low-frequency optical conductivity. The calculation of this
prominent feature in a d-wave superconductor taking into account the relevant
vertex corrections and approximate conservation laws\cite{roschPRL,roschFL} is
left as a challenge for the future -- while the $T=0$ results of this paper
can provide a reference for this calculation, the simple methods used here
will not be sufficient to describe the finite temperature regime.

{\it Acknowledgements}: We thank M. Gr\"uninger, P.~Hirschfeld, D. van der
Marel, J. Orenstein, M.~Scheffler and P. W\"olfle for useful discussions and the SFB 608
and the Emmy N\"other program of the DFG for financial support.


\end{document}